\documentclass[prx,aps,twocolumn,superscriptaddress,notitlepage,nofootinbib]{revtex4-2}
\usepackage{amssymb,epsfig,amsmath,bm,subfigure,graphicx,textcomp,url,float,color,cases}
\usepackage{epsfig}
\usepackage{textcomp}
\usepackage[normalem]{ulem}

\usepackage[colorlinks=true, urlcolor=blue, anchorcolor=blue, citecolor=blue,filecolor=blue,linkcolor=blue,menucolor=blue%,pagecolor=blue
]{hyperref}

\usepackage{color}

\usepackage[titletoc,title]{appendix}

\begin{document}
\title{Age-structured hydrodynamics of ensembles of anomalously diffusing particles with renewal resetting}

\author{Baruch Meerson}
\email{meerson@mail.huji.ac.il}
\affiliation{Racah Institute of Physics, Hebrew University of Jerusalem, Jerusalem 91904, Israel}

\author{Ohad Vilk}
\email{ohad.vilk@mail.huji.ac.il}
\affiliation{Racah Institute of Physics, Hebrew University of Jerusalem, Jerusalem 91904, Israel}

\begin{abstract}
We develop an age-structured hydrodynamic (HD) theory which describes the collective behavior of $N\gg 1$ anomalously diffusing particles under stochastic renewal resetting. The theory treats the age of a particle -- the time since its last reset -- as an explicit dynamical variable and allows for resetting rules which introduce global inter-particle correlations. The anomalous diffusion is modeled by the scaled Brownian motion (sBm): a Gaussian process with independent increments, characterized by a power-law time dependence of the diffusion coefficient, $D(t)\sim t^{2H-1}$, where $H>0$. We apply this theory to three different resetting protocols: independent resetting to the origin (model~A), resetting to the origin of the particle farthest from it (model~B), and a scaled-diffusion extension of the ``Brownian bees" model of Berestycki et al, Ann. Probab. \textbf{50}, 2133 (2022). In all these models non-equilibrium steady states are reached at long times, and we determine the steady-state densities. For model A the (normalized to unity) steady-state density coincides with the steady-state probability density of a single particle undergoing sBM with resetting to the origin. For model B, and for the scaled Brownian bees, the HD steady-state densities are markedly different: in particular, they have compact supports for all $H>0$. The age-structured HD formalism can be extended to other anomalous diffusion processes with renewal resetting protocols which introduce global inter-particle correlations.

\end{abstract}

\maketitle
\nopagebreak
\section{Introduction}

Stochastic resetting refers to the class of stochastic processes where a random process is intermittently interrupted, and the system is instantaneously returned to a prescribed configuration. 
The simplest and most studied example is a single Brownian particle on the line whose position is reset to its initial value at random times drawn from a Poisson process \cite{EM2011}. Already this minimal model exhibits key signatures of resetting: the emergence of a non-equilibrium stationary state (NESS) with nontrivial properties and an optimizable mean first-passage time to a specified target. 
Over the last decade, the resetting paradigm has been extended to a broad variety of processes and reset protocols, including non-Poissonian resetting, space-dependent reset rules, resetting in confinement, and resetting in external potentials; see the topical reviews 
\cite{Evansreview,gupta2022stochastic} and references therein. Applications span diverse contexts such as stochastic search and foraging, reaction and enzyme kinetics, population dynamics, queuing and computer algorithms, and optimization of stochastic processes subject to restart \cite{Evansreview, gupta2022stochastic, kundu2024preface, meir2025first,church2025accelerating}.

In many physical and biological systems resetting involves interacting degrees of freedom. Resetting in many-body systems with local or nonlocal interactions has been shown to produce rich collective phenomena, including nontrivial correlations and scaling regimes absent in the single-particle setting \cite{NagarGupta,Berestycki1,Berestycki2,VAM2022,SSM,vatash2025many,biroli2023extreme,biroli2025resetting}. On the other hand, particle diffusion in heterogeneous or viscoelastic media, in crowded intracellular environments, or along disordered landscapes is often anomalous \cite{jeon2011vivo,metzler2014anomalous,vilk2022ergodicity, vilk2022unravelling}, so that the mean-squared displacement (MSD) scales nonlinearly with time, $\langle x^2(t)\rangle \sim t^{2H}$, with $H\neq 1/2$. The regime $0<H<1/2$ corresponds to \emph{subdiffusion}, typically arising in crowded or viscoelastic environments, while $H>1/2$ describes \emph{superdiffusion}, associated with long-range correlations or heterogeneous energy landscapes. 

The combined effect of anomalous transport and stochastic resetting of a single particle has been explored in a variety of settings, including continuous-time random walks, scaled Brownian motion and other anomalous diffusion processes under reset \cite{biswas2025resetting,Bodrova2019,Bodrova2019a,wang2022restoring,liang2025ultraslow}. However, the collective behavior of many anomalously diffusing particles, particularly under nonlocal reset rules, remains essentially unexplored, and a systematic theoretical framework for such systems is lacking.

In this paper we develop a versatile hydrodynamic (HD) formalism which provides a leading-order description to the collective dynamics of $N\gg 1$ anomalously diffusing particles for a broad class of resetting rules, including the rules which introduce global inter-particle correlations. For concreteness and analytical tractability, we focus here on scaled Brownian motion (sBm): a Gaussian process with independent increments, characterized by a power-law time dependence of the diffusion coefficient, $D(t)\sim t^{2H-1}$, where $H$ can be any positive number \cite{Jeon,bodrova2016underdamped,Lim,Safdari}. In the presence of resetting, even a single particle performing sBm exhibits a rich behavior with two possible formulations of the model. In the first formulation -- the nonrenewal resetting -- the spatial resetting events do not affect the time-dependence of the diffusion coefficient \cite{Bodrova2019a}. Here we focus on the second formulation -- the renewal resetting -- where  the internal clock of a particle is reset along with its spatial coordinate \cite{Bodrova2019}.

The key element of our HD formalism is an explicit account for the \emph{age structure} of the population of particles, where the age of a particle, $\tau$, is defined as the time passed since its last reset. Age-structured models are widely used in demography and epidemiology \cite{webb1985theory,charlesworth1994evolution,brauer2012mathematical,inaba2017age}, but to our knowledge they have not yet been applied to many-particle stochastic resetting models.  We argue that such an extension is useful when the underlying diffusion is anomalous, because the transport properties of particles performing anomalous diffusion often depend on their history \cite{golding2006physical,weber2010bacterial,bronstein2009transient,caspi2000enhanced}.

The age-structured HD formalism for anomalously diffusing particles with resetting clearly identifies the three basic components of the dynamics: the $\tau$–dependent diffusive transport in space,  the aging in $\tau$, and the particular renewal resetting mechanism implemented through a boundary condition at $\tau=0$. This formalism circumvents a single-particle description and applies to a broad class of time-dependent renewal processes which introduce inter-particle correlations.

We illustrate the age-structured HD formulation on three representative models 
which differ among themselves by their (Poissonian) resetting rules. In model~A each particle is reset individually to the origin.  In Model~B only the particle farthest from the origin is reset to $x=0$. Finally, for the scaled Brownian bees the farthest particle is reset to the location of a particle chosen at random\footnote{\label{footnote1}{In the original formulation \cite{Berestycki1,Berestycki2} of the Brownian bees model each particle can branch into two particles and, at each branching event, the  particle farthest from the origin is removed. This model, however, is equivalent to a resetting model where the particle farthest from the origin is reset to the location of any of the particles chosen at random.}}. In each of these models a NESS emerges at long times, and we use our HD theory to determine the steady-state density fields. For model~A, the HD steady-state density coincides with the previously known steady-state single-particle distribution \cite{Bodrova2019}, as to be expected for independent resetting. For model~B and for the scaled Brownian bees, the HD steady-state densities are markedly different from the single-particle probability density. In particular, they exhibit compact support for all $H>0$. In the special case $H=1/2$ we recover the previously known HD results for resetting Brownian particles in models~A and~B \cite{VAM2022} and for the ``standard" Brownian bees \cite{Berestycki1,Berestycki2}. Throughout the paper we compare our results for the three models with Monte-Carlo simulations.

In Secs.~\ref{modelA}-\ref{sbees} we present the age-structured HD  descriptions, and determine the steady-state densities, for the three resetting models. We conclude the paper with a short discussion in Sec.~\ref{discussion}. The simulation algorithm is briefly described in the Appendix.

\section{Model A}
\label{modelA}
\subsection{Microscopic Model}

We consider \(N \gg 1\) independent particles moving on the real line. Each particle performs the sBm \cite{Lim,Jeon,Safdari} with  exponent $H>0$ and diffusion coefficient \(D\):
\begin{equation}
\label{micro}
\langle[x_i(t)-x_i(s)]^2\rangle = 2 D\,|t-s|^{2H},\qquad i=1,\dots,N.
\end{equation}
In model A, one particle is randomly chosen at random times, with total Poisson rate \(N r\), and reset to the origin. Upon resetting, the particle's clock is also reset to zero, and the particle resumes the sBm.  The single-particle ($N=1$) version of this model was introduced and analyzed by Bodrova et al. \cite{Bodrova2019}.

\subsection{$N\to \infty$: Age-structured HD}

In the HD limit $N\to\infty$ one can describe the $N$-particle system by a coarse-grained density $u(x,t)$, which we rescale by $N$, so that
\begin{equation}
\label{consA}
\int_{-\infty}^{\infty} u(x,t) \, dx = 1\,.
\end{equation}
A key element of our HD theory for the sBm is an explicit account for the individual ``clocks", or ages, of the particles. We introduce the \emph{age-structured} density $n(x,\tau,t)$ (also rescaled by $N$), where $0\leq \tau\leq t$ is the particle age, that is the time since the last reset. The coarse-grained \emph{total} density $u(x,t)$ is then obtained by integrating the age-structured density $n(x,\tau,t)$ over all ages $\tau$ between $0$ and $t$:
\begin{equation}
u(x,t) = \int_0^{t} n(x,\tau,t) \, d\tau.
\end{equation}
The HD equation for the age-structured density has the following form:
\begin{eqnarray}
\label{mainA}
\partial_t n(x,\tau,t) &+& \partial_{\tau} n(x,\tau,t) = 2H D \, \tau^{2H-1} \, \partial_x^2 n(x,\tau,t) \nonumber \\&-& r n(x,\tau,t), \quad |x|<\infty, \quad 0<\tau<t\,.
\end{eqnarray}
The $\tau$-dependent diffusion term describes the scaled diffusion. The linear decay term describes the independent resetting of particles. Finally, the advection term $\partial_{\tau} n$ describes the particle aging. In the absence of diffusion and resettings, the age distribution would follow the simple first-order equation $\partial_t n(x,\tau,t) + \partial_{\tau} n(x,\tau,t) =0$. The characteristics of this equation are described by the equation $d\tau/dt=1$, so the particles' ages 
$\tau$ increase linearly with time $t$ as to be expected.

Importantly, the instantaneous resetting of the particles to the origin enters this formulation as a boundary condition at zero age, $\tau=0$:
\begin{equation}
\label{BCtauA}
n(x,0,t) = r  \,\delta(x)\,.
\end{equation}
Finally, the particle conservation  reads
\begin{equation}
\label{consB}
\int_{-\infty}^{\infty} dx \, \int_0^{t} d\tau\, n(x,\tau,t)  \equiv \int_{-\infty}^{\infty}  u(x,t) \, dx  = 1\,.
\end{equation}
As one can see, the parameter $N$ drops out of the rescaled HD problem. 

For the standard Brownian particles, $H=1/2$, Eq.~(\ref{mainA}) simplifies to
\begin{eqnarray}
\partial_t n(x,\tau,t) + \partial_{\tau} n(x,\tau,t) &=& D \, \partial_x^2 n(x,\tau,t) - r n(x,\tau,t), \nonumber \\ \quad 0\leq \tau\leq t\,.
\end{eqnarray}
Integrating both sides of this equation over $\tau$ from $0$ to $t$ and using Eq.~(\ref{BCtauA}), we reproduce the familiar equation  \cite{VAM2022}
\begin{equation}
\label{simpleA1}
\partial_t u(x,t) = D \, \partial_x^2 u(x,t) - r u(x,t) + r \, \delta(x), 
\end{equation}
which describes model A for the standard Brownian particles.

\subsection{Steady State} \label{sec:steady_state}

\begin{figure}[ht]
\centering
\includegraphics[clip,width=0.45\textwidth]{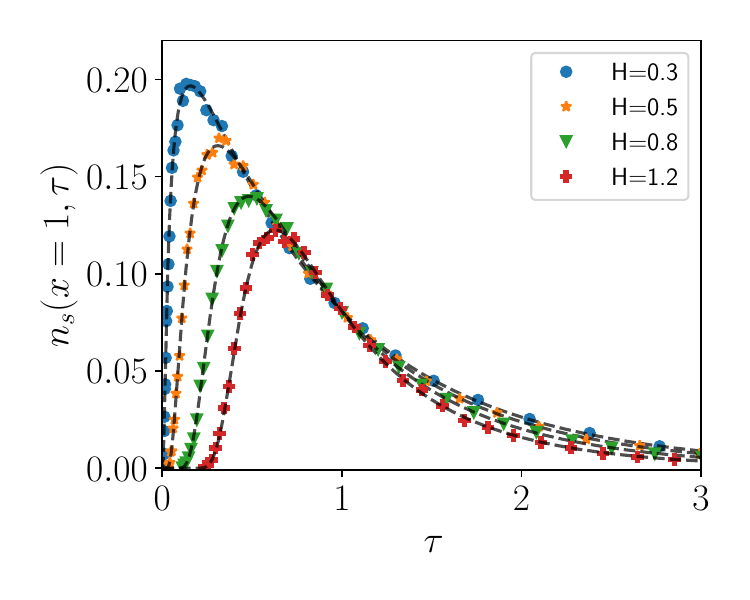}
\caption{Steady-state age-structured density $n_s(x, \tau)$ at $x=1$ versus $\tau$ for model $A$ with $r=D=1$ and four different values of $H$ (see legend). Symbols: simulations with $N = 10^5$ particles. The simulated density histograms are computed by averaging over  100 configurations  observed at different times at intervals of $\Delta t = 10$ between them. Black dashed lines: Eq. \eqref{nsA} for each of the $H$ values.}
\label{comparisonA}
\end{figure}

At $t\to \infty$ the age-structured density $n_\text{s}(x,\tau,t)$ approaches a steady state  
\begin{equation}
n_\text{s}(x,\tau) = \lim_{t\to\infty} n(x,\tau,t)
\end{equation}
with a nontrivial age structure. To determine $n_\text{s}(x,\tau)$, we drop the $t$-derivative term in Eq.~(\ref{mainA}) and obtain the equation
\begin{eqnarray}
\label{steadyeqA}
\partial_{\tau} n_\text{s}(x,\tau) &=& 2H D\,\tau^{2H-1}\,\partial_x^2 n_\text{s}(x,\tau) - r n(x,\tau,t)\,, \nonumber \\ 
&&|x|<\infty\,,\quad 0<\tau<\infty\,,
\end{eqnarray}
which should be solved with the boundary condition 
\begin{equation}\label{steadyincondA}
n_\text{s}(x,\tau=0) = r  \delta(x)\,,
\end{equation}
If we now treat $\tau$ as ``time",  Eq.~(\ref{steadyeqA})  describes an effective nonstationary (in time $\tau$) decay of particles diffusing  in the bulk following their instantaneous release at $x=0$ and $\tau=0$. The solution of this problem is straightforward, and the resulting age-structured steady-state density is
\begin{equation}\label{nsA}
n_\text{s}(x,\tau) = \frac{r \,e^{-\frac{x^2}{4 D \tau ^{2 H}}-r \tau }}{\sqrt{4\pi D}\, \tau ^{H}}\,.
\end{equation}
Figure \ref{comparisonA}  compares this simple but nontrivial result with our simulations. At $\tau=0$ the age-structured density vanishes 
identically for all $x\neq 0$. This is to be expected, because all newly reset particles are placed at the origin. At a given $x\neq 0$, $n_\text{s}(x,\tau)$ has a maximum at an age $\tau^*$ determined by the algebraic equation $Hx^2/(2D\,\tau^{*\,2H}) = r\,\tau^* + H$ which has a unique solution $\tau^*>0$.

\begin{figure*}[ht]
\centering
\includegraphics[clip,width=0.8\textwidth]{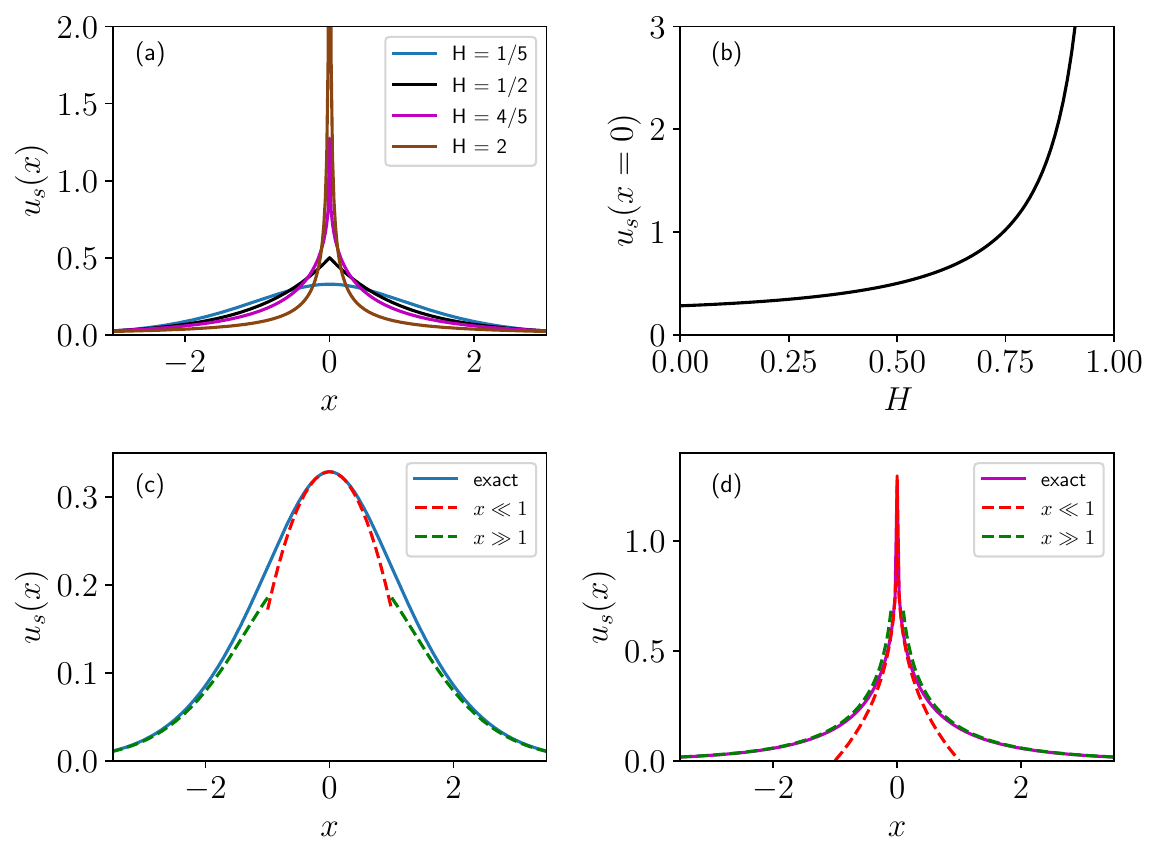}
\caption{Steady-state total density $u_{\text{s}}(x)$  for model A with $r=D=1$. (a) $u_{\text{s}}(x)$ for $H=1/5$ (blue), $1/2$ (black), $4/5$ (magenta), and $2$ (brown), see legend. 
These results were obtained by numerically evaluating the integral in Eq. \eqref{usA}.
(b) The $H$-dependence of the maximum density reached at $x=0$ for $0<H<1$. For $H>1$ the maximum density diverges, but this singularity is integrable. (c) and (d): $u_{\text{s}}(x)$, see Eq. \eqref{usA}, alongside with its large- and small-$x$ asymptotics  (\ref{largexA}) and (\ref{smallxA}), respectively, for $H=1/5$ (c) and $H=4/5$ (d).}
\label{figModelA}
\end{figure*}

The steady-state total HD density $u_\text{s}(x)$ is obtained by integrating Eq.~(\ref{nsA}) over the age variable $\tau$ from zero to $t=\infty$:
\begin{equation}\label{usA}
u_\text{s}(x) = \frac{r}{\sqrt{4\pi D}}\int_{0}^{\infty} d\tau\,\frac{\,e^{-\frac{x^2}{4 D \tau ^{2 H}}-r \tau }}{\tau ^{H}}\,.
\end{equation}
Equation~(\ref{usA}) perfectly coincides with the steady-state position distribution $p_s(x)$ of a \emph{single} particle which performs sBm with resetting. The latter expression for $p_s(x)$ was obtained, by a different method, in Ref. \cite{Bodrova2019}.  This coincidence is to be expected because of the independence of the particles in model A, and it serves as a sanity test of the age-structured HD model. As we will see shortly, the steady-state densities in two other models, that we consider here, are quite different from the single-particle density.

The characteristic spatial scale, defined by Eq.~(\ref{usA}), is $\ell_0\sim (D/r)^H$. Let us consider the $|x|\gg\ell_0$ and $|x|\ll\ell_0$ asymptotics of $u_{\text{s}}(x)$ [and of $p_\text{s}(x)$]. The $|x|\gg \ell_0$  asymptotic can be obtained by the saddle-point method (see also Ref. \cite{Bodrova2019}), and it is the following:
\begin{align}\label{largexA}
u_\text{s}(|x|\gg \ell_0)\simeq \sqrt{\frac{r}{2 (2 H+1) D}} 
\left(\frac{H}{2 Dr}\right)^{\frac{1-2 H}{2 (2 H+1)}} | x| ^{\frac{1-2 H}{2 H+1}} 
 \nonumber\\\times\exp \left[-\frac{(2 H+1) \left(\frac{H}{2 D}\right)^{\frac{1}{2 H+1}}
   r^{\frac{2 H}{2 H+1}} | x| ^{\frac{2}{2 H+1}}}{2 H}\right]\,.
\end{align}
The $|x|\ll \ell_0$ asymptotic can be obtained via a truncated Taylor expansion of the integrand at small $|x|$, leading to
\begin{align}
\label{smallxA}
&u_\text{s}(|x| \!\ll \! \ell_0) \!\simeq\! \frac{1}{\sqrt{4\pi}}\times \nonumber\\&
\begin{cases}\Gamma(1\!-\!H) \!- \!\Gamma(1\!-\!3H)\frac{x^2}{4} ,&  \text{$0<H<1/3$},\\ 
\Gamma(1\!-\!H)\!+\! 
\frac{1}{2H}\Gamma\left(\frac{1}{2}\!-\!\frac{1}{2H} \right) \left(\frac{x^2}{4}\right)^{\frac{1-H}{2H}},& \text{ $1/3<H<1$},\\
\frac{1}{2H}\Gamma\left(\frac{1}{2}\!-\!\frac{1}{2H}\right)\left(\frac{x^2}{4}\right)^{\frac{1-H}{2H}}, & \text{$H>1$}\,,
\end{cases}
\end{align}
where we have set $r=D=1$ for brevity. As one can see from Eq.~(\ref{smallxA}), the maximum density is observed at $x=0$, and it is finite only for $H<1$. For $H\geq 1$ it diverges, but this singularity is integrable and therefore legitimate. For $H = 1/2$ Eq.~(\ref{usA}) yields $u_\text{s}(x) = (1/2)\,e^{-|x|}$ in agreement with  Ref. \cite{VAM2022}.

\begin{figure}[ht]
\centering
\includegraphics[clip,width=0.45\textwidth]{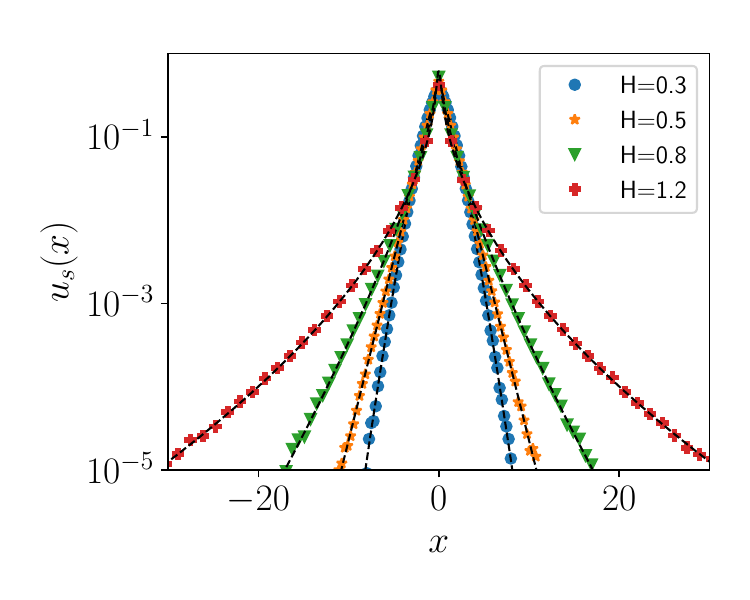}
\caption{Steady-state total density $u_s(x)$ for model A with $r=D=1$ and four values of $H$ (see legend). Symbols: simulations with $N = 10^5$ particles. The simulated density histograms are computed by averaging over  100 configurations  observed at different times at intervals of $\Delta t = 10$ between them. Black dashed lines: Eq. \eqref{usA} for each of the $H$ values.}
\label{u_modelA_sim}
\end{figure}

Figure \ref{figModelA} shows the steady-state total density $u_{\text{s}}(x)$ for several values of  $H$, and the maximum steady-state density at $x=0$ as a function of $H$. The bottom panels compares the exact densities with their large- and small-$x$ asymptotics  (\ref{largexA}) and (\ref{smallxA}).  Finally, Fig. \ref{u_modelA_sim} compares the steady state density $u_{\text{s}}(x)$ for different $H$ with simulations.

According to Eq. (\ref{nsA}), the steady-state HD density lives on the whole line $|x|<\infty$.  However, the actual average system radius $\bar{\ell}$, defined as the average distance to the origin of the farthest particle, is of course finite for any finite $N$. Formally speaking, this effect is missed by a HD theory, but for model A it is in fact captured, with logarithmic accuracy, by the simple order-of-magnitude estimate $N \int_{\bar{\ell}}^{\infty} u_{\text{s}}(x) dx \simeq 1$ which has an obvious physical meaning. Using the large-$|x|$ asymptotic (\ref{largexA}) of $p_{\text{s}}(x)\equiv u_{\text{s}}(x)$, we obtain in the leading order:
\begin{equation}\label{ell}
\bar{\ell} (H) \simeq 2^{H+1} \left(\frac{H}{ 2H +1}\right)^H \frac{\left(\ln N\right)^{H+\frac{1}{2}}}{\sqrt{2 H+1}} \,.
\end{equation}
This expression agrees with the leading term of the large-$N$ asymptotic of the exact microscopic expression for $\bar{\ell}$ \cite{vilk2025collective}. For $H=1/2$ Eq.~(\ref{ell}) predicts
\begin{equation}\label{ellhalf}
\bar{\ell} (H=1/2) \simeq \ln N \,,
\end{equation}
in agreement with Ref. \cite{VAM2022}.

\section{Model B}
\label{modelB}
\subsection{Microscopic Model}
The only difference, but an important one,  between models A and  B is that in model B it is the particle farthest from the origin which is reset to the origin. Clearly, this resetting rule introduces non-local inter-particle correlations.
As a result, the collective behavior of particles in model B is not reducible to a single-particle behavior, but in the limit of $N\to \infty$, is amenable to a HD theory. 

\subsection{$N\to \infty$: Age-Structured HD}

A key feature of the HD particle density  $u(x,t)$ (which we again rescale by $N$) in model B is that it has a compact support, $|x|<L(t)$. In the particular case of $H=1/2$ this property was uncovered in Ref. \cite{VAM2022}.  The edges of support, $|x|=L(t)$, act as effective absorbing walls. Their a priori unknown position,  at all times,  is dictated implicitly by the particle conservation: 
\begin{equation}
\label{cons}
\int_{-L(t)}^{L(t)} u(x,t) \, dx = 1, \qquad u(|x|=L(t),t) = 0,
\end{equation}
so one has to deal with a moving-boundary problem. We again introduce the age-structured density $n(x,\tau,t)$, where $0\leq \tau\leq t$ is the time since the last resetting, and $u(x,t) = \int_0^{t} n(x,\tau,t) \, d\tau$. The HD equation for model B is, in the dimensional units,
\begin{eqnarray}
\label{mainB}
\partial_t n(x,\tau,t) + \partial_{\tau} n(x,\tau,t) &=& 2H D \, \tau^{2H-1} \, \partial_x^2 n(x,\tau,t), \nonumber \\  0<\tau<t,&&  |x|<L(t)\,.
\end{eqnarray}
Note the key structural difference: in model A the independent resetting produces the bulk loss term $-rn$ see Eq.~(\ref{mainA}), whereas in model B the farthest-particle resetting is encoded in the absorbing boundaries at $|x|=L(t)$, with no bulk loss term.

Like in model A, the resetting of particles to the origin, accompanied by the age renewal, is described by the boundary condition 
\begin{equation}
n(x,0,t) = r  \delta(x)\,.
\end{equation}
Finally, the particle conservation takes the form
\begin{equation}
\label{consB}
\int_{-L(t)}^{L(t)} dx \, \int_0^{t} d\tau\, n(x,\tau,t)  \equiv \int_{-L(t)}^{L(t)}  u(x,t)  dx  = 1\,.
\end{equation}
For the standard Brownian motion, $H=1/2$, Eq.~(\ref{mainB}) simplifies to
\begin{eqnarray}
\partial_t n(x,\tau,t) + \partial_{\tau} n(x,\tau,t) &&= D \, \partial_x^2 n(x,\tau,t), \nonumber \\ 
0\leq \tau\leq t, && |x|<L(t)\,.
\end{eqnarray}
Integrating this equation over $\tau$, we obtain
\begin{align}
\label{simpleB1}
\partial_t u(x,t) = D \, \partial_x^2 u(x,t) + r \, \delta(x), \nonumber \\ \quad |x|<L(t), \quad v(\pm L(t),t)=0, 
\end{align}
and 
\begin{equation}
\label{simpleB2}
\int_{-L(t)}^{L(t)} u(x,t) \, dx = 1\,.
\end{equation}
Equations~(\ref{simpleB1}) and (\ref{simpleB2}) coincide with model B for the standard Brownian particles \cite{VAM2022}.

\begin{figure}[t]
\centering
\includegraphics[clip,width=0.45\textwidth]{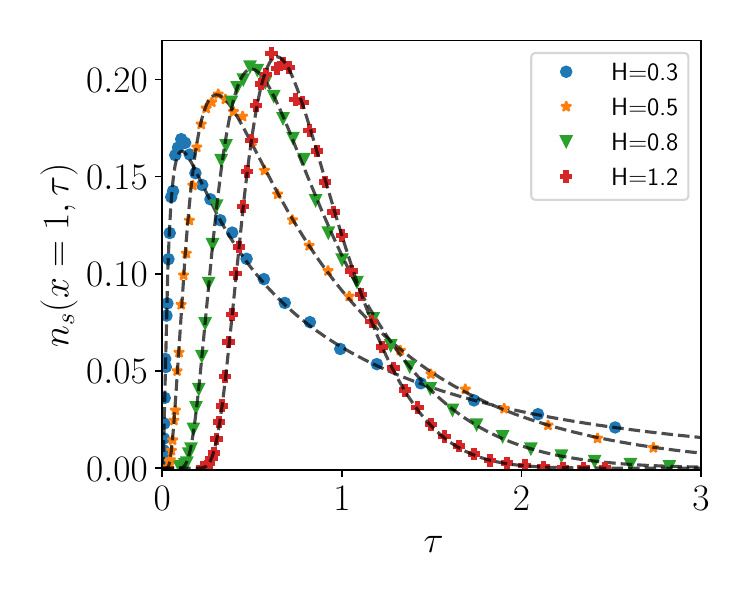}
\caption{Steady-state age-structured density $n_s(x, \tau)$ at $x=1$ for model $B$ with $r=D=1$ and four different values of $H$ (see legend).  Symbols: simulations with $N = 10^5$, where the simulated density histograms are computed by averaging over 100 configurations observed at different times at intervals of $\Delta t = 10$ between them. Black dashed lines: Eq. \eqref{nsB} for each of the $H$ values.} 
\label{comparisonB}
\end{figure}

\begin{figure*}[ht]
\centering
\includegraphics[clip,width=0.75\textwidth]{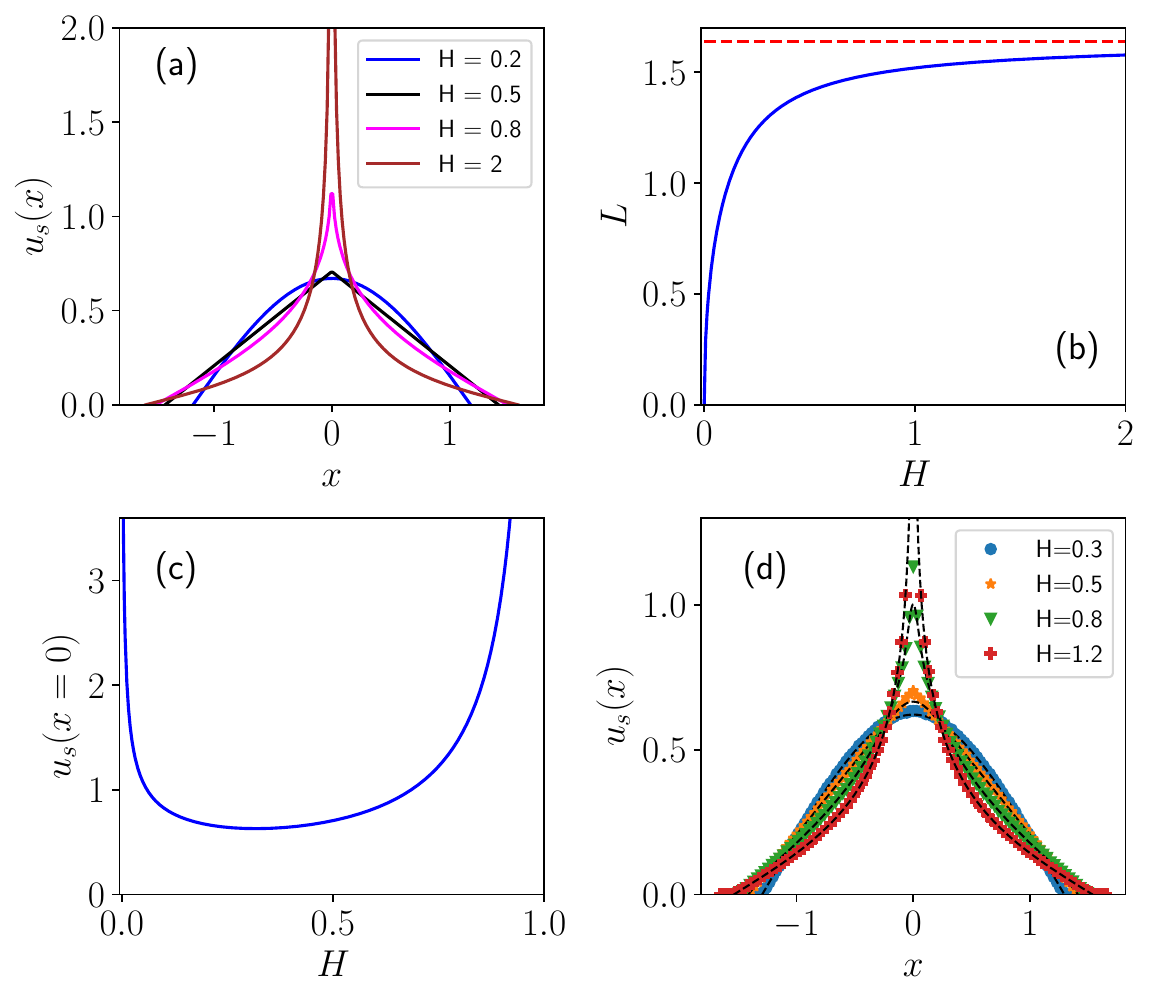}
\caption{Steady-state density $u_{\text{s}}(x)$  for model B with $r=D=1$.  (a)  $u_{\text{s}}(x)$, see Eq. \eqref{usB}, for $H=1/5$ (blue), $1/2$ (black), $4/5$ (magenta), and $2$ (brown). (b) The support radius $L=L(H)$ as described by Eq. \eqref{LvsH} (blue solid line) and its limit for $H\to \infty$, Eq. \eqref{limitL} (red dashed line). (c) The $H$-dependence of the maximum density reached at $x=0$ for $0<H<1$. For $H>1$ the maximum density diverges, but this singularity is integrable. (d) Comparison of the theoretically predicted $u_{\text{s}}(x)$, Eq. \eqref{usB} (black dashed lines), with simulations (symbols) with $N = 10^5$ particles and four different values of $H$ (see legend). The density histograms are computed via averaging over 100 configurations of the system, observed at intervals of $\Delta t = 10$.}
\label{figModelB}
\end{figure*}

\subsection{Steady State}

As $t\to \infty$, the absorbing walls at $|x|=L(t)$ come to rest. The steady-state value of $L$ reflects a balance between the effective particle source at the origin and the diffusion flux of the particles into the absorbing walls.
The age-structured density $n_\text{s}(x,\tau,t)$ approaches a steady state 
\begin{equation}
n_\text{s}(x,\tau) = \lim_{t\to\infty} n(x,\tau,t)
\end{equation}
with a nontrivial age structure. The steady state is described by the equation
\begin{equation}
\label{steadyeqB}
\partial_{\tau} n_\text{s}(x,\tau) = 2H D\,\tau^{2H-1}\,\partial_x^2 n_\text{s}(x,\tau), \quad |x|<L =\text{const},
\end{equation}
with the boundary conditions $n_\text{s}(\pm L,\tau)=0$, and the ``initial" condition 
\begin{equation}\label{steadyincondB}
n_\text{s}(x,\tau=0) = r  \delta(x), 
\end{equation}
where we again treat $\tau$ as time. With these conditions, Eq.~(\ref{steadyeqB})  describes an effective nonstationary (in time $\tau$) diffusion flux, created by an instantaneous release of mass at $x=0$,  to the absorbing boundaries at $x=\pm L$, where the parameter $L$ is a priori unknown.  This
problem can be easily solved by introducing a ``stretched time" $\tau^{2H}$ and expanding the solution over the eigenmodes $\sim \cos\left[(2m+1)\pi x /2L\right]$. The resulting age-structured steady-state solution reads 
\begin{eqnarray}\label{nsB}
n_\text{s}(x,\tau) = \frac{r}{L} \sum _{m=0}^{\infty } e^{-\frac{\pi ^2 (2 m+1)^2 D \tau ^{2
   H}}{4 L^2}} &&\cos \left[\frac{(2 m+1) \pi  x}{2 L}\right]\,, \nonumber 
   \\ \quad |x|<L\,,
\end{eqnarray}
and $0$ for $|x|>L$. Figure \ref{comparisonB} compares this result with our simulations.  
At $\tau=0$ the age-structured density vanishes 
identically for all $0<|x|<L$, because all newly reset particles are placed at the origin.  At a given $0<|x|<L$, the density (\ref{nsB}) as a function of $\tau$ rises from zero, peaks at a finite $\tau^*$, and then decays. Unlike in model A, here there is no effective particle loss in the bulk. As a result,  the peak value of $n_s$ increases with $H$ -- in contrast to model~A (compare Figs.~\ref{comparisonB} and~\ref{comparisonA}).

The normalized particle density is obtained by integrating Eq.~(\ref{nsB}) over the age variable $\tau$:
\begin{align} \label{usB} 
u_\text{s}(x) =&\int_0^\infty n_\text{s}(x,\tau)\, d\tau \nonumber
\\  =& \left(\frac{2}{\pi }\right)^{1/H} \Gamma \left(1+\frac{1}{2
   H}\right) \left(\frac{L^2}{D}\right)^{\frac{1}{2 H}}  \nonumber \\ & \times \frac{r}{L}  \sum
   _{m=0}^{\infty } \frac{\cos \left[\frac{(2m+1) \pi 
   x}{2 L}\right]}{ (2 m+1)^{\frac{1}{H}}}\,.
\end{align}
What is left is to determine $L=L(H)$ from the conservation law, see Eq.~(\ref{consB}), and we obtain
\begin{align}\label{LvsH}
L(H)=2 \pi ^{1\!+\!H} \sqrt{D} \bigg\{r \Gamma \left(1\!+\!\frac{1}{2 H}\right)
   \left[\zeta \left(1\!+\!\frac{1}{H},\frac{1}{4}\right)\right. \nonumber\\  \left.-\zeta
   \left(1\!+\!\frac{1}{H},\frac{3}{4}\right)\right]\bigg\}^{-H}\,, 
\end{align}
where $\zeta(s,a)$ is the generalized Riemann zeta function \cite{Wolfram}.  As $H\to \infty$,  $L(H)$ approaches a finite limit. In the units $r=D=1$ it is equal to
\begin{equation}
\label{limitL}
L(H\to \infty) = \frac{32\, e^{-\gamma /2} \,\Gamma \left(5/4\right)^4}{\pi ^2} = 1.6398109\dots\,,
\end{equation}
where $\gamma$ is  Euler's constant. 

The maximum value of the density \eqref{usB} is observed at the origin. Similarly to model A, it is finite for $H<1$ and diverges for $H\geq 1$, but this singularity is integrable and therefore legitimate. Interestingly, for model B, the  maximum density as a function of $H$ reaches its minimum at an intermediate value of $0<H<1$ which depends on the reset rate $r$ but is independent of $D$.

Figure \ref{figModelB} shows the steady-state total densities $u_{\text{s}}(x)$ for several values of $H$, the support radius $L=L(H)$ and the maximum steady-state density at $x=0$ as a function of $H$.

For $H = 1/2$ Eqs.~(\ref{usB}) and~(\ref{LvsH}) simplify to
\begin{equation}
u(x) = \frac{r}{2D} (L - |x|), \qquad |x| < L, \qquad L = \sqrt{2D/r},
\end{equation}
in agreement with Ref.~\cite{VAM2022}.

\section{Scaled Brownian bees}
\label{sbees}

\subsection{Microscopic Model}
This model is a natural extension of the Brownian bees model proposed by Berestycki et al. \cite{Berestycki1,Berestycki2} and studied in Refs. \cite{Berestycki1,Berestycki2,MS2021,SSM,SVS}. In the original formulation of the Brownian bees model  $N\gg 1$ particles perform independent Brownian motions and independently branch at a given rate but, at each branching event, the  particle farthest from the origin is removed. This model can be easily reformulated in terms of resetting, where the particle farthest from the origin is reset to the location of any of the particles chosen at random. Similarly to model B, this resetting rule introduces global inter-particle correlations, so that the collective behavior is irreducible to the single-particle behavior. In this case as well, the HD theory proves very useful.

\subsection{$N\to \infty$: Age-structured HD}

As in the original Brownian bees model \cite{Berestycki1,Berestycki2}, the coarse-grained density $u(x,t)$ of the scaled Brownian bees has compact support $|x|<L(t)$ in the limit of $N\to \infty$.  Here too there are effective absorbing walls at $|x|=L(t)$, and their position is implicitly governed by the particle conservation, see Eq.~(\ref{consB}).

Our HD formulation of this model is also age-structured and employs the density $n(x,\tau,t)$. The latter obeys the equation  
\begin{eqnarray}
\label{mainbees}
\partial_t n(x,\tau,t) + \partial_{\tau} n(x,\tau,t) &=& 2H D \, \tau^{2H-1} \, \partial_x^2 n(x,\tau,t), \nonumber \\ 
0<\tau<t, &&
|x|<L(t)\,,
\end{eqnarray}
which is identical to its counterpart (\ref{mainB}) for model B.   The particle conservation  also
has the same form (\ref{consB}) as in model B. Differently from models A and B, the boundary condition at $\tau=0$ is now nonlocal in $\tau$:
\begin{equation}
\label{BCbees}
  n(x,\tau=0,t) = r\, u(x,t) \equiv\int_{0}^{t}   n(x,\tau,t) \,d\tau\,.
\end{equation}
This equation reflects the fact that the ``newborn"  particles, which appear at reset rate $r$, can be accommodated at locations of particles of \emph{all} ages. 

One can check that, for $H=1/2$, Eqs.~(\ref{mainbees}) and (\ref{BCbees}) can be reduced to the equations  for the total density $u(x,t)$ of the original Brownian bees model \cite{Berestycki1,Berestycki2}.

\subsection{Steady State}

\begin{figure*}[t]
\centering
\includegraphics[clip,width=0.8\textwidth]{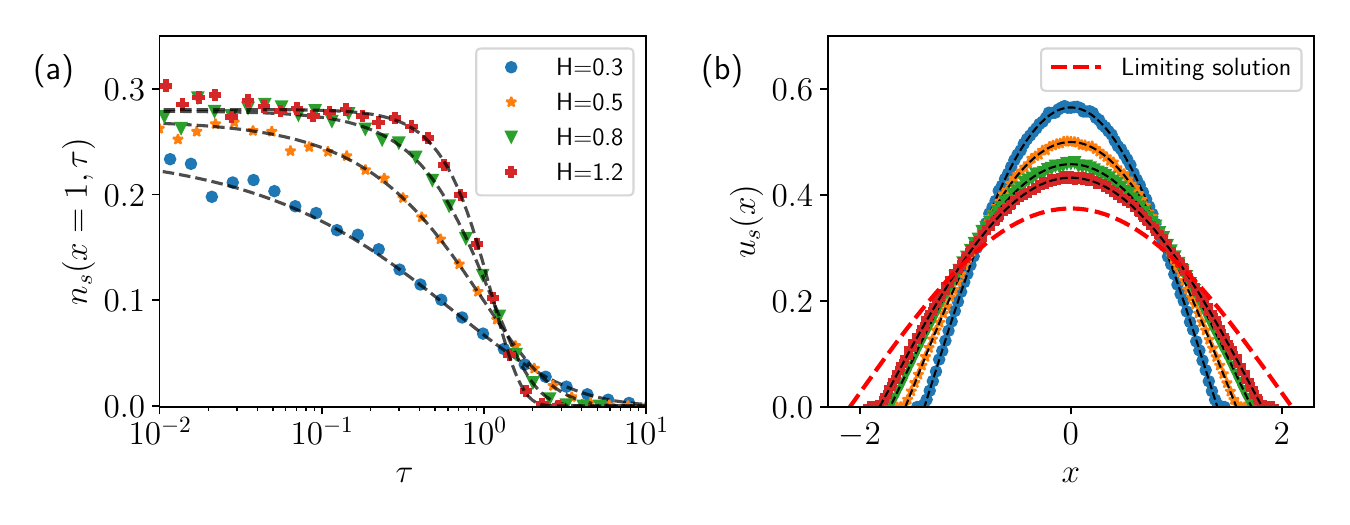}
\caption{Scaled Brownian bees with $r=D=1$  and four different values of $H$ (see legend). (a) Steady state age-structured density $n_s(x, \tau)$ at $x=1$. Simulations (symbols) are compared to Eq. \eqref{nsbees} (black dashed lines). (b) The steady state total density $u_s(x)$. Here Eq. \eqref{ubees} (black dashed lines) is compared with simulations (symbols). Also shown is the limiting solution~(\ref{limitprofile}) obtained at $H\to \infty$ (red dashed line). In both panels $N = 10^5$, and simulated density histograms are computed by averaging over  100 configurations  observed at different times at intervals of $\Delta t = 10$ between them.}
\label{comparisonbees}
\end{figure*}

Here too, at $t\to \infty$, the age-structured density $n_\text{s}(x,\tau,t)$ approaches a steady state 
with a nontrivial age structure. The steady state density $n_s(x,\tau)$ is described by the same equation (\ref{steadyeqB}) as in model B.
The boundary conditions are again $n_\text{s}(\pm L,\tau)=0$, but the initial condition (in terms of the time $\tau$)  is different. This condition follows from Eq.~(\ref{BCbees}) and has the form
\begin{equation}
\label{steadyincondbees}
n_{\text{s}}(x,\tau=0) = \int_{0}^{\infty}   n_{\text{s}}(x,\tau) \,d\tau\,,
\end{equation}
where we have again switched to the dimensionless units $r=D=1$. In spite of the unusual form of the ``initial" condition~(\ref{steadyincondbees}), the steady-state problem can be solved quite easily. Indeed, since the age-structured density $n_{\text{s}}(x,\tau)$ must be positive for all $|x|<L$, only the lowest Dirichlet eigenmode of the operator $d^2/dx^2$ can contribute. This immediately leads to the following ansatz:
\begin{equation}
\label{nsbees}
    n_{\text{s}}(x,\tau) = A  \exp \left(-\frac{\pi^2 \tau^{2H}}{4L^2}\right)\,\cos \left(\frac{\pi x}{2L}\right)\,. 
\end{equation}
The a priori unknown system radius $L=L(H)$  and  the amplitude $A=A(H)$ can be determined from Eqs.~(\ref{consB}) and~(\ref{steadyincondbees}). After a straightforward algebra we obtain
\begin{align}
\label{LAbees}
     &L(H)= \frac{\pi}{2}\left[\Gamma\left(1+\frac{1}{2H}\right)\right]^{-H}, \nonumber \\  &A(H)=\frac{1}{2} \left[\Gamma \left(1+\frac{1}{2 H}\right)\right]^H\,.
\end{align}
Fig. \ref{comparisonbees}(a) compares the theoretically predicted age-structured steady-state density $n_s(x,\tau)$ with simulations. Notice that, in contrast to model B, the age-structured density in the present case has a maximum at the origin, as the particles can be reset to any $|x|\leq L$.

Integrating Eq.~(\ref{nsbees}) over $\tau$ and using Eq.~(\ref{LAbees}), we obtain the steady-state total density of the scaled Brownian bees:
\begin{eqnarray}
\label{ubees}
 u_{\text{s}}(x) = \frac{1}{2}\, \left[\Gamma \left(1+\frac{1}{2 H}\right)\right]^H &\cos \left[\Gamma \left(1+\frac{1}{2
   H}\right)^H\,x \right]\,, \nonumber \\
   |x|<L(H)\,,
\end{eqnarray}
and zero for $|x|>L(H)$. For $H=1/2$,  Eq.~(\ref{ubees}) reduces to that for the Brownian bees, $u_s(x)=(1/2) \cos x$, and $L=\pi/2$ \cite{Berestycki1,Berestycki2}. As $H \to 0$, $L(H)$ goes to zero as $\sim H^{1/4}$, while the maximum density at $x=0$ goes to infinity as $H^{-1/4}$. 

Like in model B, the support radius $L(H)$ reaches a finite limit of as $H\to \infty$: $L(H \to \infty)= (\pi/2)\, e^{\gamma/2}$.
Remarkably, $u_{\text{s}}(x)$ approaches a universal limiting solution in this limit: 
\begin{equation}
\label{limitprofile}
u_{\text{s}}(x, H\to \infty) = \frac{1}{2} e^{-\frac{\gamma}{2}} \cos \left(e^{-\frac{\gamma}{2}} x\right)\,,\quad |x|<\frac{\pi}{2} e^{\gamma/2} \,.
\end{equation}
This solution is shown in Fig. \ref{comparisonbees}(b) alongside with a comparison of the predicted steady-state total density~(\ref{ubees}) for different values of $H$ with simulations.

\section{Discussion}
\label{discussion}

We have presented a versatile age-structured HD formalism which captures the leading-order $N\to \infty$ behavior 
of large ensembles of anomalously diffusing particles subject to a variety of resetting protocols, including protocols which
introduce strong non-local inter-particle correlations. We have illustrated the formalism by considering
scaled Brownian particles under three different resetting protocols. In all three models we have determined
the age-structured particle density at the steady state, which then yields the steady-state total particle density. 
We argue that the age-structured formulation is quite general. Beyond the applicability to different resetting rules, the approach can accommodate other underlying transport processes whose propagators are explicitly time-dependent. In particular, it can be used when the scaled Brownian motion is replaced by a continuous-time random walk, where an aged-structured description, without resets, has already been suggested \cite{berry2016quantitative}.

As we have shown, the age-structured HD formalism provides a valuable qualitative and quantitative insight into the physics of these systems. However, by construction, this formalism ignores fluctuations. These can be studied, in a straightforward manner, in model A because of its non-interacting character \cite{vilk2025collective}.  However, for correlated ensembles, exemplified by Model B and by scaled Brownian bees, such a straightforward analysis, based on the single-particle picture, is not feasible.  Yet the fluctuations in these globally correlated systems are undoubtedly of great interest. One interesting statistics is the system's radius $\ell$ where even in the standard case $H=1/2$ a nontrivial $\ln N/N$ scaling behavior of the variance of $\ell$ with $N$ was uncovered \cite{VAM2022,SSM}. A natural framework for studying this and other statistical properties of the correlated models is fluctuating hydrodynamics -- a mesoscopic theory in which the HD description is augmented by suitable Langevin-type noise terms that capture, on large spatiotemporal scales, the randomness of the underlying microscopic processes. Developing a first-principle aged-structured fluctuating-hydrodynamic framework for the sBm with resetting under different resetting rules therefore appears to be a  promising direction for future research.

\bigskip
\noindent
{\bf Acknowledgment}.  B. M.  was supported by the Israel Science Foundation 
(Grant No. 1579/25).

\bibliography{references}% Produces the bibliography via BibTeX.

\appendix

\section{Monte Carlo simulations}

We simulate $N$ particles undergoing sBm with exponent $H$ and the model-specific random resets. Between resetting events, the position increments of a particle $i$ are Gaussian with variance 
$2 D_{\text{eff}, i} \,dt$, where $t_i$ is the internal clock of particle $i$, that is the time since the last reset for that particle. See Ref. \cite{LimMuniandy2002} for a detailed explanation on how to simulate sBm. 

As explained in the main text, the resets occur with exponential rate $r N$. At each reset only one particle is reset. For model A we choose a particle at random and set its coordinate to 0; for model B we find the particle with the largest absolute value of the coordinate and set it to 0; and for the  scaled Brownian bees we find the particle with the largest absolute value of the coordinate and place it at the location of a randomly selected particle \cite{SSM,VAM2022}. 

The characteristic relaxation time of each of these three models is $\mathcal{O}(1/r)$ in macroscopic units of time, i. e.,  the system “forgets” its initial condition after $\mathcal{O}(N)$ resetting events. Consequently, a simulation run up to time $10^5$ provides 
many effectively uncorrelated samples,
which can be used for testing the theory.

The simulations were done using Python version 3.11 with NumPy for numerics and Matplotlib for visualization. 

\end{document}